\documentclass[conference,a4paper]{IEEEtran}
\usepackage{psfig,amsmath,amssymb,amsthm}
\usepackage{subfigure,cite,graphicx,psfig,amsmath,amssymb,eufrak,mathrsfs,epsf,epsfig}
\usepackage[active]{srcltx}
\begin{document}
\newcommand{\efield}{\textbf{E}}
\newcommand{\hfield}{\textbf{H}}
\newcommand{\ecurr}{\textbf{J}}
\newcommand{\curl}{\nabla\times}
\newcommand{\diver}{\nabla\cdot}
\newcommand{\pilot}{\overrightarrow{\textbf{r}}}
\newcommand{\unitr}{\hat{\textbf{r}}}
\newcommand{\bfr}{\textbf{r}}
\newcommand{\bfE}{\textbf{E}}
\newcommand{\hankfn}{h_{n}^{(1)}}
\newcommand{\hankfnp}{h_{n'}^{(1)}}
\newcommand{\resist}{R}
\newcommand{\power}{P}
\newcommand{\noise}{N_{W}}
\newcommand{\x}{\textbf{X}}
\newcommand{\diag}{\textup{diag}}
\newcommand{\bfJ}{\textbf{J}}
\newcommand{\bfs}{\textbf{s}}
\newcommand{\bfe}{\textbf{e}}
\newcommand{\bfu}{\textbf{u}}
\newcommand{\bfT}{\textbf{T}}
\newcommand{\bfU}{\textbf{U}}
\newcommand{\bfY}{\textbf{Y}}
\newcommand{\bfX}{\textbf{X}}
\newcommand{\bfI}{\textbf{I}}
\newcommand{\bftau}{\boldsymbol{\tau}}
\newcommand{\bfphi}{\boldsymbol{\phi}}
\newcommand{\bfPhi}{\boldsymbol{\Phi}}
\newcommand{\bftheta}{\boldsymbol{\theta}}
\newcommand{\bfzeta}{\boldsymbol{\zeta}}
\newcommand{\bfSigma}{\boldsymbol{\Sigma}}
\newcommand{\condition}{\textbf{K}_{\tilde{\x}}:\textup{tr}(\textbf{K}_{\tilde{\x}})\le\power}
\newcounter{casenum}
\newcommand{\cn}{\refstepcounter{casenum} \thecasenum}
\theoremstyle{definition}
\newtheorem{thm}{Theorem}
\newtheorem{prop}{Proposition}
\newtheorem{definition}{Definition}
\newtheorem{corollary}{Corollary}
\newtheorem{remark}{Remark}

\sloppy

\title{The Capacity of Wireless Channels:\\ A Physical Approach}
\author{
  \IEEEauthorblockN{Wonseok Jeon}
  \IEEEauthorblockA{Department of EE, KAIST\\
    Daejeon, Korea\\
    Email: wonsjeon@kaist.ac.kr}
  \and
  \IEEEauthorblockN{Sae-Young Chung}
  \IEEEauthorblockA{Department of EE, KAIST\\
    Daejeon, Korea\\
    Email: sychung@ee.kaist.ac.kr}
}
\maketitle
\begin{abstract}
In this paper, the capacity of wireless channels is characterized based
on electromagnetic and antenna theories with only minimal assumptions.
We assume the transmitter can generate an arbitrary current distribution inside 
a spherical region and the receive antennas are uniformly distributed on a bigger sphere surrounding the transmitter.
The capacity is shown to be $(\alpha P/N_0) \log e$ [bits/sec] in the limit of large number of receive antennas, where $P$ is the transmit power constraint, $\alpha$ is the normalized density of the receive antennas and $N_0$ is the noise power spectral density. Although this result may look trivial, it is surprising in two ways. First, this result holds regardless of the bandwidth (bandwidth can even be negligibly small). Second, this result shows that the capacity is irrespective of the size of the region containing the transmitter. This is against some previous results that claimed the maximum degrees of freedom is proportional to the surface area containing the transmitter normalized by the square of the wavelength.
Our result has important practical implications since it shows that even a compact antenna array with negligible bandwidth and antenna spacing well below the wavelength can provide a huge throughput as if the array was big enough so that the antenna spacing is on the order of the wavelength.

\end{abstract}
\section{Introduction}
The capacity of multiple-input multiple-output (MIMO) channels has been considered to be severely limited by the size of antenna arrays. The capacity scaling for a three-dimensional network~\cite{Franceschettietal:2009} and the degree-of-freedom analysis for a polarimetric antenna array~\cite{PoonTse:2011} were provided based on the spherical vector wave decomposition, and both showed the number of usable channels is proportional to the surface area enclosing the transmitter.

There have been a lot of attempts to squeeze more antennas into a given space. For example, polarimetric antennas were used to achieve degrees-of-freedom gains by using the wave polarization~\cite{PoonTse:2011},\cite{Andrewsetal:2001},~\cite{Chaeetal:2011}. Also, the MIMO cube was introduced, which consists of twelve dipoles located on the edges of the cube to increase the degree of freedom in a limited space~\cite{Gustafssonetal:2006},~\cite{Getuetal:2005}. Besides, it was recently shown that even when the antenna spacing at the receiver is negligibly small, two degrees of freedom can be achieved for a two-user multiple access channel, which was previously thought impossible~\cite{Ivr:2011}.

In this paper, we attempt to characterize the ultimate limit of wireless communication by deriving the capacity of wireless channels with only minimal assumptions. We assume the transmit antennas are confined inside a sphere
of a certain radius but otherwise completely arbitrary and assume the total transmit power is constrained to be $P$. We also assume arbitrarily many receive antennas so that they do not become a bottleneck. We show the capacity is given by $(\alpha P/N_0) \log e$ irrespective of the bandwidth, where $P$ is the transmit power constraint, $\alpha$ is the normalized density of the receive antennas and $N_0$ is the noise power spectral density. Interestingly, the capacity is irrespective of the size of the source region, which is due to the availability of arbitrarily many spectral channels of equal quality. This result is in contrast with the previous work in~\cite{PoonTse:2011} that claimed that the maximum degrees of freedom is proportional to the surface area of the region containing the transmitter.
Our result has important practical implications since it shows that even a compact antenna array with negligible bandwidth and antenna spacing well below the wavelength can provide a huge throughput as if the array was big enough so that the antenna spacing is on the order of the wavelength.


\emph{Notation}: $(\cdot)^*, (\cdot)^t$ and $(\cdot)^\dagger$ are vector complex conjugate, matrix transpose, and matrix conjugate-transpose. $j_n^{(1)}(\cdot)$ and $h_n^{(1)}(\cdot)$ denote the spherical Bessel function of the first and the third kind, respectively, $Y_{nm}(\cdot,\cdot)$ is the spherical harmonic function. The unit vectors on the spherical coordinate are $\hat{\bfr},\hat{\boldsymbol{\theta}}$ and $\hat{\boldsymbol{\phi}}$. $\textbf{I}_{n}$ is the $ n \times n $ identity matrix.

\section{System Model} \label{sec:systemmodel}

\begin{figure}[!t]
\centering
\includegraphics[angle=0,width=0.4\textwidth]{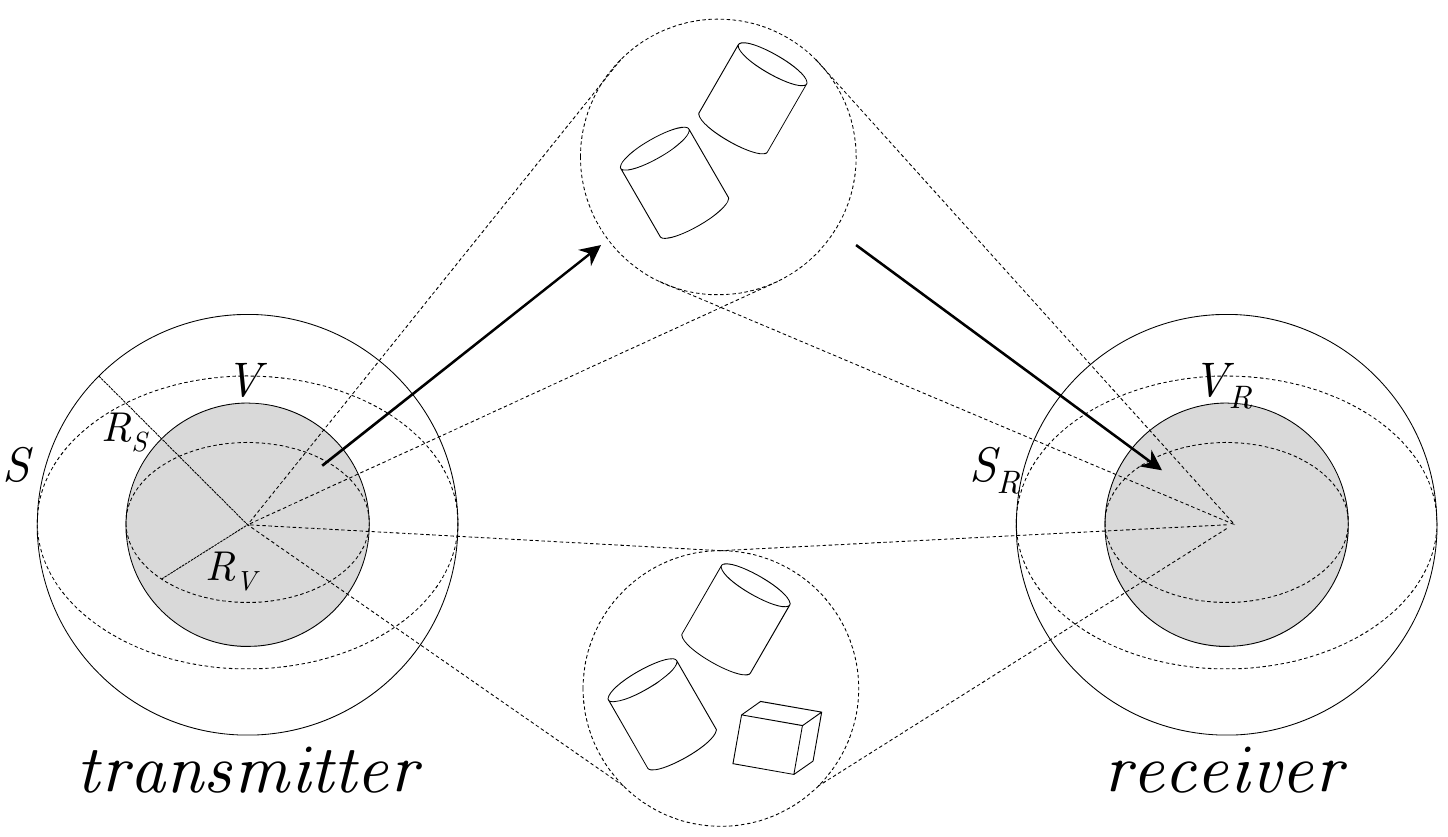}
\caption{Illustration on scattering environment.}
\label{fig:fig_dipoles}
\end{figure}
\begin{figure}[!t]
\centering
\subfigure[]{
\includegraphics[angle=0,width=0.2\textwidth]{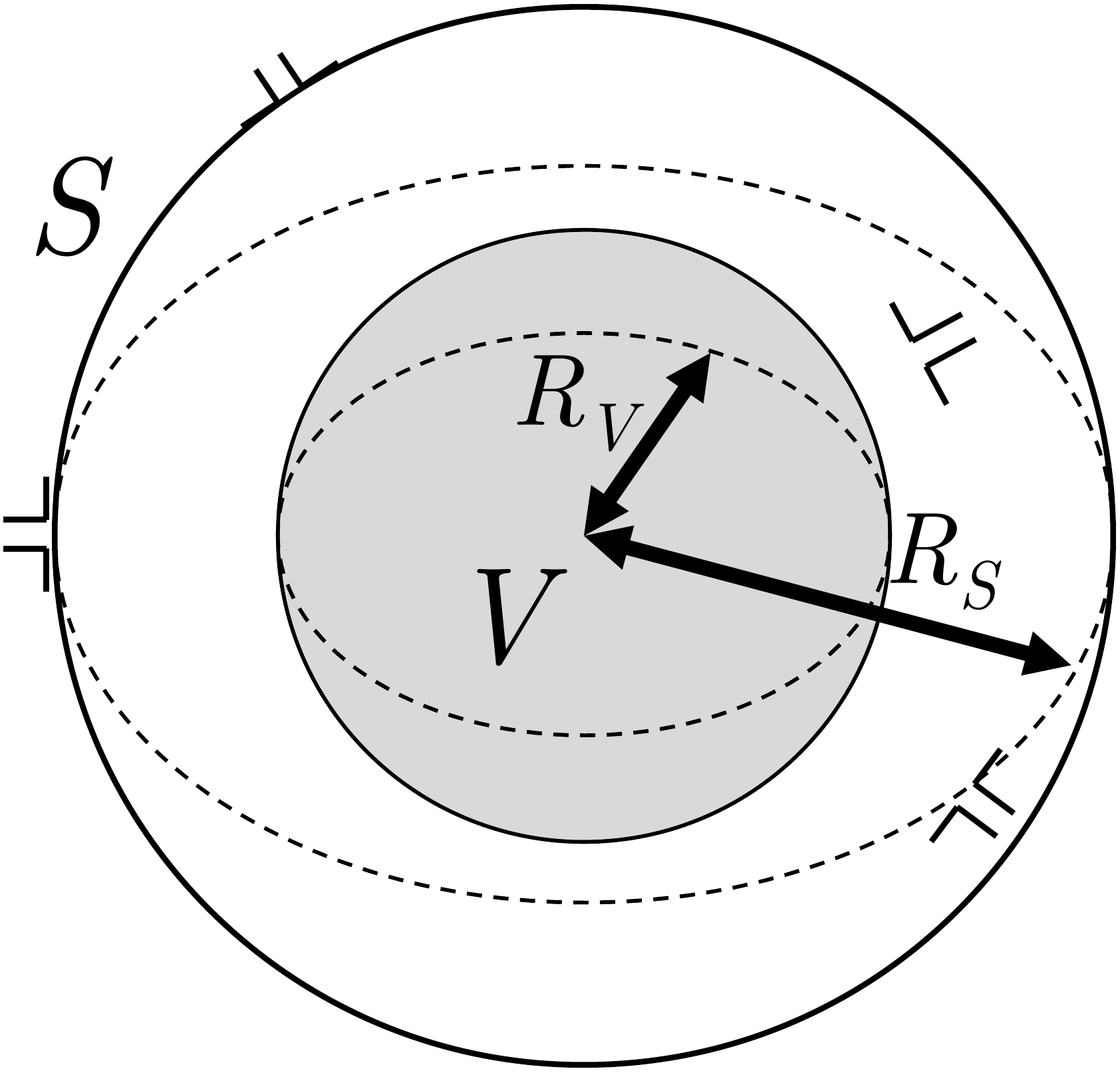}
\label{fig:fig_dipole}
}
\subfigure[]{
\includegraphics[angle=0,width=0.2	\textwidth]{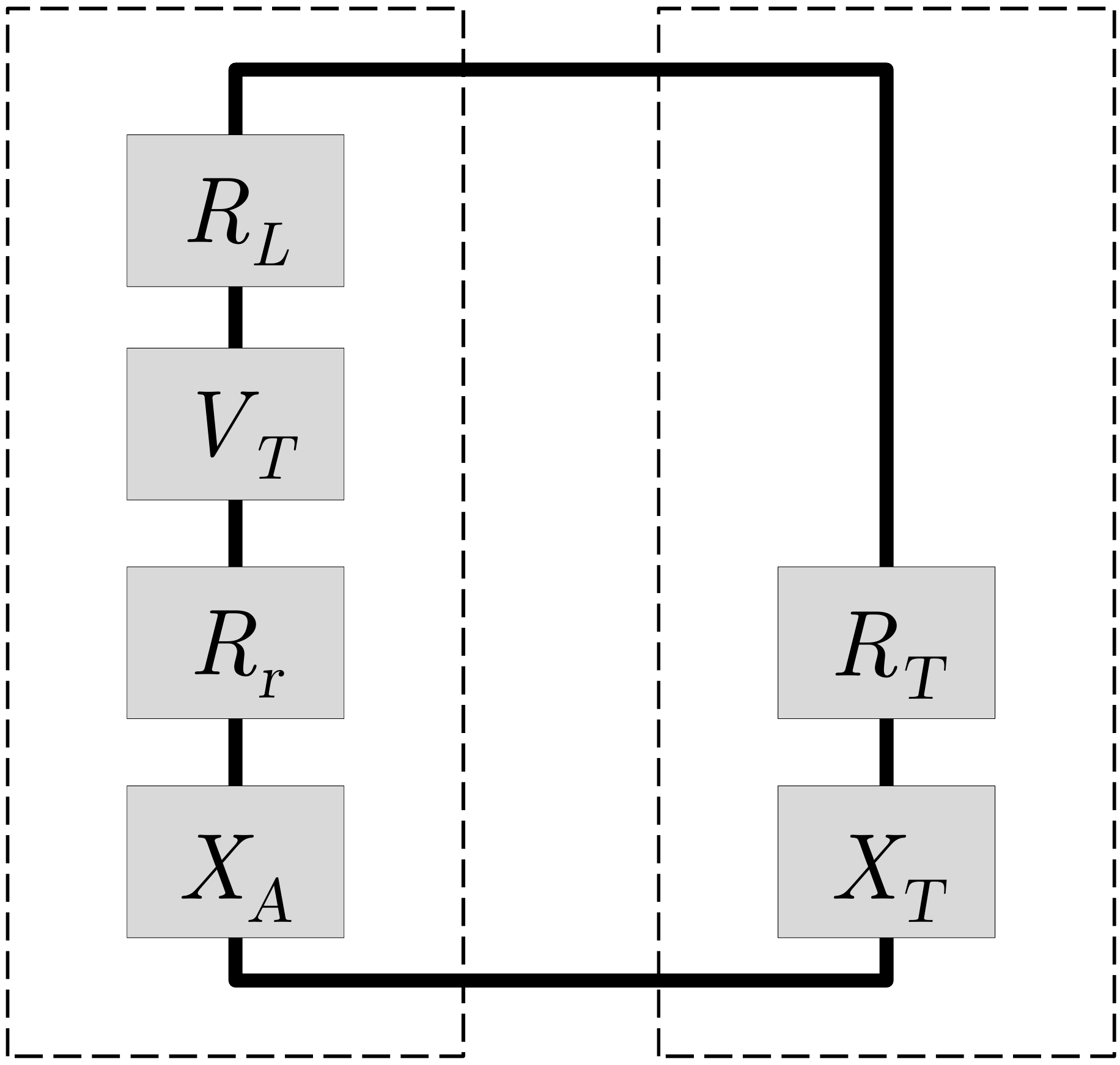}
\label{fig:fig_thevenin}
}
\caption{(a) Channel model with the transmitter inside $V$ and receiving short dipole antennas uniformly distributed on $S$ (b) Equivalent Thevenin circuit for a receiver. The left half models the antenna while the right half is the load.}
\end{figure}

Let us first consider the channel model depicted in Fig.~\ref{fig:fig_dipoles}, where a transmitter is inside a spherical region $V$ with radius $R_V$ and a receiver is inside another spherical region $V_R$. The overall channel can be decomposed into three parts, the channel from the transmitter in $V$ to the $\mathbf{E}$ field on the surface $S$ with radius $R_S$ surrounding the transmitter, the scattering environment from $S$ to $S_R$, another surface surrounding the receiver, and finally the last part from $S_R$ to the receiver in $V_R$. In our paper, we focus on the first part because the last part can be analyzed similarly as the first one using reciprocity and the second channel is simply a scattering environment. More specifically, we assume receive antennas (short dipole antennas) are located uniformly on $S$ as shown in Fig.~\ref{fig:fig_dipole} and characterize the capacity of the channel from the source in $V$ to the receive antennas on $S$. We assume both $V$ and $S$ are centered at the origin.

Assume $\bfJ(\bfr)$ is the current density at $\bfr\in V$ due to the transmitter in $V$, and $\bfE(\bfr)$ is the electric field at $\bfr \in \mathbb{E}^3$ generated from the current. The Green function $\textbf{G}(\bfr ,\bfr ')$ relates the current to the electric field as
\begin{align}
\efield(\bfr )=i\omega\mu\int_{V}\textbf{G}(\bfr ,\bfr ')
\ecurr (\bfr ')d\bfr ',\label{eq:E10}
\end{align}
where $\omega$ is the carrier frequency, and $\mu$ is the permeability of $V^{C}$~\cite[p.376]{Chew}. Both $\bfJ(\bfr)$ and $\bfE(\bfr)$ are complex baseband representations. In this paper, we omit time indices for simplicity.

We assume the following transmit power constraint:
\begin{align}
E\left[
-\mathfrak{Re}\left\{i\omega\mu\int_V \int_{V}\left(\textbf{G}(\bfr ,\bfr ')
\ecurr (\bfr ')\right)\cdot\bfJ^*(\bfr)d\bfr ' d\bfr\right\}\right]\le P.
\label{eq:pc}
\end{align}
%
%

Let $\mathbf{s}_q\in S, q=1,...,N,$ denote the location of the $q$-th receive dipole, where the locations are uniformly distributed on $S$. Formally, we use the following definition for uniformity.
\begin{definition}
A set of points $\bfs_{q,N}\in S, q=1,...,N$, $N\in\mathbb{Z}^+$ is said to be \emph{uniformly distributed on $S$} with respect to a function $f$ if
\begin{align*}
\lim_{N\rightarrow\infty}\frac{4\pi}{N}\sum_{q=1}^{N}f(\bfs_{q,N})=\int_Sf(\bfr)d\Omega.
\end{align*}
We denote $\bfs_{q,N}$ as $\bfs_q$ for simplicity. 
\end{definition}
Throughout this paper, we assume the set of points $\bfs_q,q=1,...,N$ is uniformly distributed w.r.t. $(\bfu_p^*(\bfr)\cdot \boldsymbol{\hat{r}}) (\bfu_{p'}(\bfr)\cdot \boldsymbol{\hat{r}})$, $(\bfu_p^*(\bfr)\cdot \boldsymbol{\hat{\theta}}) (\bfu_{p'}(\bfr)\cdot \boldsymbol{\hat{\theta}})$, and $(\bfu_p^*(\bfr)\cdot \boldsymbol{\hat{\phi}}) (\bfu_{p'}(\bfr)\cdot \boldsymbol{\hat{\phi}})$ for each $p,p'\in\mathbb{Z}^+$, where $\bfu_{p}(\bfr)$ is defined in \ref{sec:SVD}. It is easy to construct $\mathbf{s}_q$'s explicitly to satisfy the uniformity condition.

Let $\alpha=N/(8k^2R_S^2)$, $0<\alpha<1$, denote the normalized density of receive antennas, where $k=\frac{\omega}{c}$ is the wave number and $c$ is the speed of light. We will show in \textbf{Remark \ref{Remark2}} that $\alpha P$ is equal to the total received power under our assumption that the receive antennas are uniformly distributed. We assume $\alpha \ll 1$ since in practice the received power is very small compared to the transmit power. We also assume the receive antennas are sufficiently separated, i.e., the minimum distance between any two antennas is on the order of $\frac{1}{\sqrt{\alpha}k}$. Note that the mutual coupling among receive antennas and the mutual coupling between the transmitter and the receive antennas can be ignored under the assumptions.

Let $\textbf{e}_{q}$ denote the orientation of the $q$-th dipole, which is assumed to be given by
\begin{align*}
\textbf{e}_{q}=
\begin{cases}
\boldsymbol{\hat{r}},&q=1,4,7,...,\\
\boldsymbol{\hat{\theta}},&q=2,5,8,...,\\
\boldsymbol{\hat{\phi}},&q=3,6,9,...,
\end{cases}
\end{align*}
such that the antenna directions are uniform.
The received signal of the $q$-th dipole is 
\begin{align}
Y_q=\int_V \bfJ(\bfr)\cdot\bfzeta_q(\bfr)d\bfr+Z_q, q=1,...,N,\label{eq:channel1}
\end{align}
where $\bfzeta_q(\bfr)$ models the channel from $\bfJ(\bfr)$ to the received signal of the $q$-th dipole. We assume $Z_q\sim\mathcal{CN}\left(0,N_{0}W\right),q=1,...,N,$ are independent and identically distributed circularly symmetric complex Gaussian noises independent of the source signal, where $N_0$ is the noise power spectral density and $W$ is the channel bandwidth satisfying $W\ll \frac{\omega}{2\pi}$.

\section{Capacity}
\label{sec:cap}
In this section, we derive the capacity of the channel (\ref{eq:channel1}). We consider the limit of large number of receive antennas for simplifying analysis. Specifically, let $C_N$ denote the capacity of the channel \eqref{eq:channel1} satisfying the power constraint (\ref{eq:pc}). Our goal is to find the capacity $C$ of the channel in the limit $N\rightarrow\infty$ while keeping the normalized density of receive antennas $\alpha=N/(8k^2R_S^2)$ constant, i.e.,
\begin{align}
C=\lim_{N\rightarrow\infty}C_N.\label{eq:capacity}
\end{align}

Now we state our main theorem.
\begin{thm}\label{Thm2}
For a given $\alpha$, $C$ is given as
\begin{align*}
C=\frac{\alpha\power}{N_0}\log e. 
\end{align*}
The proof is shown in Subsection \ref{sec:Proposition1} using the following concepts.
\end{thm}

\subsection{Singular value decomposition of the Green function}\label{sec:SVD}
 Consider the following orthogonal bases
\begin{align*}
\textbf{U}_{nm1}(\bfr )
&=\curl\bfr \hankfn (kr)Y_{nm}(\theta,\phi),\\
\textbf{U}_{nm2}(\bfr )
&=\frac{1}{k} \curl\curl\bfr \hankfn (kr)Y_{nm}(\theta,\phi),\\
\textbf{V}_{nm1}(\bfr )
&=\curl\bfr j_{n}(kr)Y_{nm}(\theta,\phi),\\
\textbf{V}_{nm2}(\bfr )
&=\frac{1}{k} \curl\curl\bfr j_{n}(kr)Y_{nm}(\theta,\phi),
\end{align*}
for integers $n\in\mathbb{Z}^+, -n\le m\le n, l=1,2$, where $\left\{\textbf{U}_{nml}(\bfr)\right\}$ spans the vector field on $S$ and $\left\{\textbf{V}_{nml}(\bfr)\right\}$ spans the vector field in $V$. For notational simplicity, we use $p\triangleq2(n(n+1)+m-1)+l$ to represent $(n,m,l)$ such that there is one-to-one correspondence between $(n,m,l)$ and $p$.

In \cite{PoonTse:2011}, the Green function in \eqref{eq:E10} is decomposed into
\begin{align}
\textbf{G}(\bfr ,\bfr ')=ik\sum_{p=1}^{\infty}\sigma_{p}
\textbf{u}_{p}(\bfr )\textbf{v}_{p}^\dagger(\bfr '),\bfr\in S,\bfr'\in V,\label{eq:E16}
\end{align}
where $\sigma_p,p\in\mathbb{Z}^+$ are singular values and
\begin{align}
\textbf{u}_{p}(\bfr )
&=\frac{\textbf{U}_{p}(\bfr )}
{\sqrt{\int_{S}| \textbf{U}_{p}(\bfr)|^{2} d\Omega}}\triangleq\frac{1}{C_{p}}\textbf{U}_{p}(\bfr ),
\label{eq:E5}
\\
\textbf{v}_{p}(\bfr )
&=\frac{\textbf{V}_{p}(\bfr )}
{\sqrt{\int_{V}| \textbf{V}_{p}(\bfr )|^{2} d\bfr }}.\nonumber
\end{align}
Also, the current due to the transmitter can be decomposed into
\begin{align}
\ecurr (\bfr )=\sum_{p=1}^{\infty} J_{p} \textbf{v}_{p}(\bfr ),\label{eq:current}
\end{align}
where
\begin{align}
J_{p}\triangleq\int_{V} \ecurr (\bfr )\cdot \textbf{v}_{p}^*(\bfr )d\bfr, p\in\mathbb{Z}^+.\label{eq:E17}
\end{align}
Using \eqref{eq:E16} and \eqref{eq:E17}, \eqref{eq:E10} gives the electric field on $S$ as
\begin{align}
\notag\efield(\bfr )
&=-\omega\mu k
\sum_{p=1}^{\infty}\sum_{p'=1}^{\infty}\sigma_{p}J_{p'}\textbf{u}_{p}(\bfr )\int_{V}\textbf{v}_{p}^{*}(\bfr ')\cdot \textbf{v}_{p'}(\bfr ')d\bfr '\nonumber\\
&=-\eta k^{2}
\sum_{p=1}^{\infty}\sigma_{p}J_{p}\textbf{u}_{p}(\bfr ),\bfr\in S,\label{eq:E18}
\end{align}
by applying the orthogonality of $\textbf{v}_{p}(\bfr)$ and $\omega\mu=\eta k$, where $\eta=120\pi$ is the wave impedance.

\subsection{Transmit power and radiation resistance}
In this subsection, we express the transmit power given in the left hand side of (\ref{eq:pc}) in terms of $J_p$'s and define the radiation resistance for each mode.

\begin{figure}[!t]
\centering
\includegraphics[angle=0,width=0.3\textwidth]{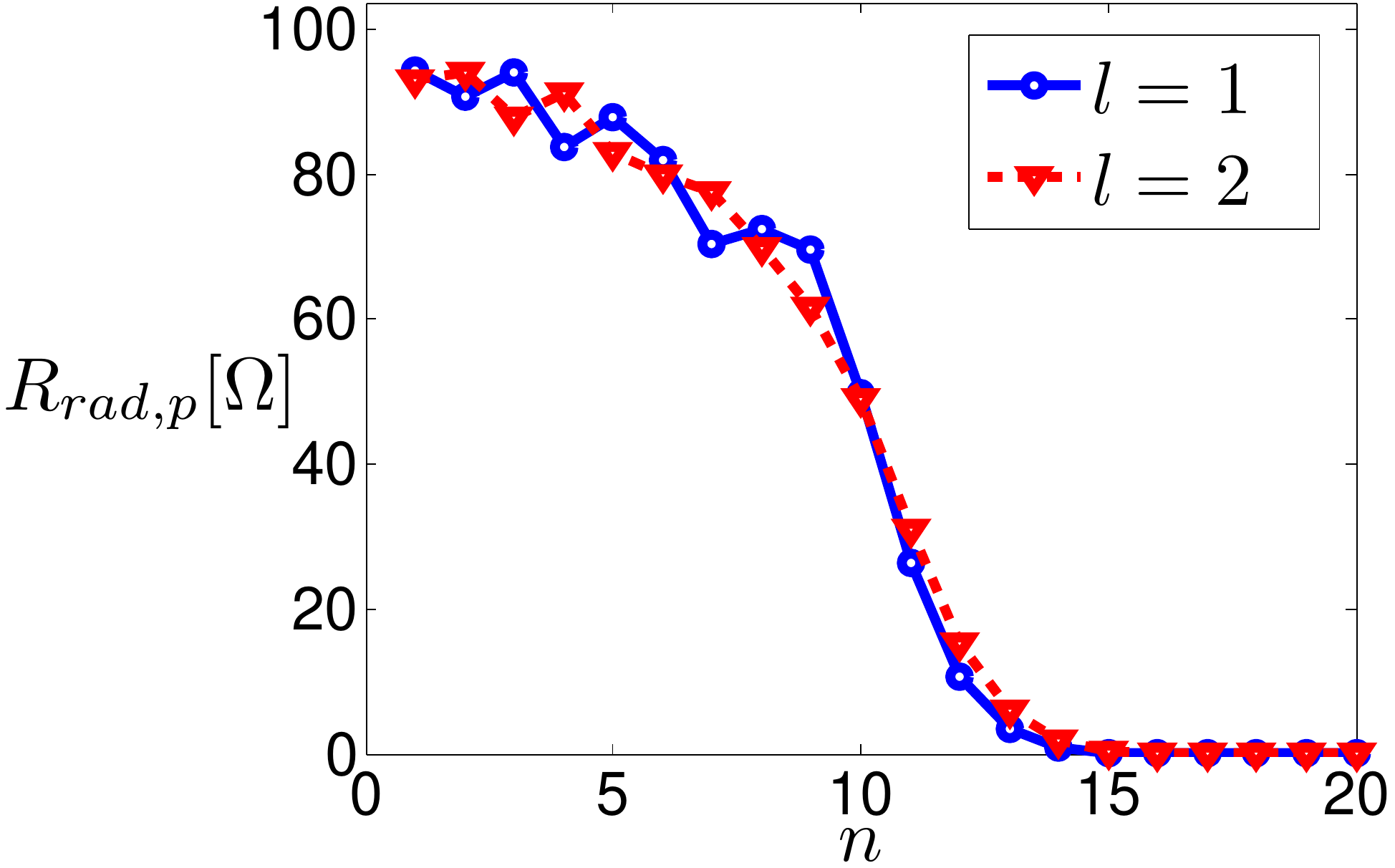}
\caption{Radiation resistance over modes when $R_V=2\lambda$.}
\label{fig:radiationresistance}
\end{figure}

\begin{prop}\label{Proposition1}
The transmit power given in (\ref{eq:pc}) is equal to the following:
\begin{align*}
E\left[\sum_{p=1}^{\infty}T_p|J_{p}|^2\right]
\end{align*}
where 
\begin{align*}
T_p
\begin{cases}
=&\frac{\eta k^2R_V^3}{4}[j_{n-1}^2(kR_V)+j_n^2(kR_V)\\
&-\frac{2n+1}{kR_V}j_{n-1}(kR_V)j_n(kR_V)], p=1,3,...,\\
=&\frac{\eta k^2 R_V^3}{4}\left\{\frac{n+1}{2n+1}[j_{n-2}^2(kR_V)+j_{n-1}^2(kR_V)\right.\\
&-\frac{2n-1}{kR_V}j_{n-2}(kR_V)j_{n-1}(kR_V)]\\
&+\frac{n}{2n+1}[j_n^2(kR_V)+j_{n+1}^2(kR_V)\\
&\left.-\frac{2n+3}{kR_V}j_n(kR_V)j_{n+1}(kR_V)]\right\},p=2,4,...,
\end{cases}
\end{align*}
and $n$ is the corresponding index for $p$. Note that $T_p$ is irrelevant to $R_S$ for all $p$. The proof is in Appendix \ref{sec:AppA}.
\end{prop}
\begin{definition}\label{sec:defi3}
The \emph{radiation resistance} $R_{rad,p}$ for each decomposed channel is defined as
\begin{align*}
R_{rad,p}=\frac{T_p}{R_V}\left[\textup{Ohm}\right], p\in \mathbb{Z}^+.
\end{align*}
Note that the radiation resistance depends only on $kR_V$ and vanishes as $n$ increases as shown in Fig.~\ref{fig:radiationresistance}.
\end{definition}
\begin{remark}
The radiated power for mode $p\in\mathbb{Z}^+$ is given by $P_p=E\left[R_{rad,p}\left|J_p\right|^{2}R_V\right]$.
\end{remark}

\subsection{Receiver model}
Each short dipole can be modeled by an equivalent circuit as depicted in Fig. \ref{fig:fig_thevenin}, where $V_{T}$ is the voltage coming from the incident wave, $\resist_{r}$ is the radiation resistance of the dipole, $\resist_{L}$ is the loss resistance, $\resist_{T}$ is the resistance of the load, $\textup{X}_{A}$ is the antenna reactance, and $\textup{X}_{T}$ is the reactance of the load~\cite[p.84]{Balanis}.
We choose $\resist_{L}=0, \textup{X}_{A}=-\textup{X}_{T}$ and $\resist_{r}+\resist_{L}=\resist_{T}$ for conjugate matching to deliver the maximum power to the load. Then, the power transmitted to the load resistor is
\begin{align*}
P_{T}=\frac{|V_{T}|^{2}}{8\resist_{T}}.
\end{align*}
We define the receive signal $Y_q$ by the $q$-th dipole as
\begin{align}
Y_{q}=\frac{V_{Tq}}{\sqrt{8\resist_{Tq}}}, q=1,...,N, \label{eq:e26}
\end{align}
so that its square is the same as the power transmitted to the load of the $q$-th receiver. In addition,
\begin{align}
\resist_{T_{q}}&=\resist_{r_{q}}=80\left(\frac{\pi L}{\lambda}\right)^{2},q=1,...,N,\label{eq:e28}
\end{align}
where $\lambda$ is the wavelength and $L$ is the length of the dipole. 
Assuming the incident field is a \emph{plane wave}~\cite[p.91]{Balanis}\footnote{This will hold asymptotically as $N\rightarrow\infty$ in our case since $R_S$ also tends to infinity as $N\rightarrow\infty$. Thus, this assumption is valid for evaluating $C$ in (\ref{eq:capacity}) as will be shown in Subsection \ref{sec:Proposition1}.}, we have 
\begin{align}
V_{Tq}=L\efield(\textbf{s}_{q})\cdot\textbf{e}_{q},q=1,...,N.\label{eq:e29}
\end{align}
Including the thermal noise across $\resist_T$, the received signal can be rewritten as
\begin{align}
Y_{q}
&=\frac{\lambda}{\sqrt{640}\pi}\efield(\textbf{s}_{q})\cdot\textbf{e}_{q}+Z_{q},q=1,...,N. \label{eq:eee}
\end{align}

Now, assume we use modes from $p=1$ to $p=M$. Then, \eqref{eq:E18} and \eqref{eq:eee} lead to
\begin{align}
Y_{q}=-\frac{\eta k}{\sqrt{160}}\sum_{p=1}^{M}
\left(
\textbf{u}_{p}(\textbf{s}_{q})\cdot\textbf{e}_{q}
\right)\sigma_{p}J_{p}+Z_{q},\label{eq:receive}
\end{align}
for each $q$. 
Let $\x=\left[ J_{1},...,J_{M} \right]^{t}$ and $\textbf{Y}=\left[Y_{1},...,Y_{N}\right]^{t}$ denote the channel input and output, respectively. Then, \eqref{eq:receive} becomes
\begin{align}
\textbf{Y}
&=g\boldsymbol{\Phi}\boldsymbol{\Sigma}\x+\textbf{Z}\label{eq:e36},
\end{align}
where $g\triangleq-\frac{\eta k}{\sqrt{160}}$, $\boldsymbol{\Sigma}=\textup{diag}\left\{\sigma_1,...,\sigma_M\right\}$, $\boldsymbol{\Phi}=\left[\boldsymbol{\phi}_{1},...,\boldsymbol{\phi}_{M}\right]$ such that
$\boldsymbol{\phi}_{p}=
\left[
\textbf{u}_{p}(\textbf{s}_{1})\cdot\textbf{e}_{1},
...,
\textbf{u}_{p}(\textbf{s}_{N})\cdot\textbf{e}_{N}
\right]^{t},p=1,...,M,$ and $\textbf{Z}=\left[Z_{1},...,Z_{N}\right]^{t}$.

\subsection{Proof}\label{sec:Proposition1}
For given $N$ and $M$, let $C_{NM}$ denote the capacity of the channel \eqref{eq:channel1} when $\ecurr (\bfr )$ has the form (\ref{eq:current}) with $J_p=0$ for all $p> M$.
Let $\bar{C}_{NM}$ denote the capacity of the same channel under the plane wave assumption at receive antennas, i.e., (\ref{eq:e36}). Then, we have
\begin{align}
C
&=\lim_{N\rightarrow\infty}C_N
=\lim_{N\rightarrow\infty}\lim_{M\rightarrow\infty}C_{NM}
\overset{(a)}{=}\lim_{M\rightarrow\infty}\lim_{N\rightarrow\infty}C_{NM}\nonumber\\
&\overset{(b)}{=}\lim_{M\rightarrow\infty}\lim_{N\rightarrow\infty}\bar{C}_{NM}.\label{eq:proofofcap}
\end{align}
Here, the limits can be exchanged in (a) since $C_{NM}$ is the capacity of a MIMO channel given by \eqref{eq:channel1} combined with \eqref{eq:current}.
Also, (b) holds since the plane wave assumption holds asymptotically as $N\rightarrow\infty$.

Before we evaluate \eqref{eq:proofofcap}, first observe that
\begin{align}
\lim_{N\rightarrow\infty}\frac{1}{N}\boldsymbol{\Phi}^\dagger\boldsymbol{\Phi}=\frac{1}{12\pi}\textbf{I}_M \label{eq:prop1}
\end{align}
for any $M$. This follows because for all $p,p'=1,...,M$ we have
\begin{align*}
\frac{12\pi}{N}\boldsymbol{\phi}_{p}^{\dagger}\boldsymbol{\phi}_{p'}
&=\frac{12\pi}{N}\sum_{q=1}^{N}\left\{\textbf{u}_{p}(\textbf{s}_{q})\cdot \textbf{e}_{q}\right\}^{*}\left\{\textbf{u}_{p'}(\textbf{s}_{q})\cdot\textbf{e}_{q}\right\}\rightarrow\delta_{pp'}
\end{align*}
as $N\rightarrow\infty$ by using \eqref{eq:E5} and uniform distribution of dipoles. 

In addition, let $\bfT=\textup{diag}\left\{T_1,...,T_M\right\}$. Then, we have
\begin{align}
\bftau\triangleq R_S\bfT^{-\frac{1}{2}}\bfSigma\rightarrow\sqrt{\frac{2}{\eta k^4}}\bfI_M \label{eq:prop2}
\end{align}
as $N\rightarrow\infty$ by using
\begin{align*}
\frac{\eta k^4R_S^2\sigma_p^2}{2T_p}=
\begin{cases}
1+\mathcal{O}\left(\frac{n^{2}}{kR_{S}}\right), \text{ for odd } p,\\
1+\frac{n+1}{2n+1}\mathcal{O}\left(\frac{(n-1)^{2}}{kR_{S}}\right)+\frac{n}{2n+1}\mathcal{O}\left(\frac{(n+1)^{2}}{kR_{S}}\right), \\
\text{ for even } p,
\end{cases}
\end{align*}
from the expansion in \cite[p.925]{Gradshteyn}: 
\begin{align*}
h_n^{(1)}(kR_S)
&=(-i)^{n+1}\frac{e^{ikR_S}}{kR_S}\left[1+\mathcal{O}\left(\frac{n^{2}}{kR_{S}}\right)\right].
\end{align*}

Now, assume $\tilde{\x}\triangleq\textbf{T}^{\frac{1}{2}}\x$ and $N_{W}\triangleq N_{0}W$. Then, 
\begin{align*}
\bar{C}_{NM}&=\max_{\condition}
W\log\left|\textbf{I} _{N}
+\frac{g^2}{\noise}
\boldsymbol{\Phi}\bfSigma\textbf{T}^{-\frac{1}{2}}
\textbf{K}_{\tilde{\x}}
\textbf{T}^{-\frac{1}{2}}\bfSigma\boldsymbol{\Phi}^{\dagger}
\right|\\
&=\max_{\condition}
W\log\left|\textbf{I} _{M}
+\frac{g^2}{\noise}
\textbf{T}^{-\frac{1}{2}}\bfSigma\boldsymbol{\Phi}^{\dagger}
\boldsymbol{\Phi}\bfSigma\textbf{T}^{-\frac{1}{2}}
\textbf{K}_{\tilde{\x}}
\right|\\
&=\max_{\condition}
W\log\left|\textbf{I} _{M}
+\frac{8\alpha g^2 k^2}{\noise}
\bftau
\left(\frac{1}{N}\boldsymbol{\Phi}^{\dagger}\boldsymbol{\Phi}\right)
\bftau
\textbf{K}_{\tilde{\x}}
\right|\nonumber\\
&\rightarrow
\max_{\condition}
W\log\left|\textbf{I}_M+\frac{\alpha}{\noise}
\textbf{K}_{\tilde{\x}}\right|\nonumber\\
&=MW
\log\left(1+\frac{\alpha P}{MN_W}\right)
\end{align*}
as $N\rightarrow\infty$, which follows by using $\textbf{K}_{\tilde{\x}}=\textbf{T}^{\frac{1}{2}}\textbf{K}_{\x}\textbf{T}^{\frac{1}{2}}$
and the property $\left|\textbf{I} _{N}+\textbf{AB}\right|=\left|\textbf{I}_{M}+\textbf{BA}\right|$ and applying \eqref{eq:prop1} and \eqref{eq:prop2} for convergence. Finally, we get
\begin{align*}
C=\lim_{M\rightarrow\infty}MW\log\left(1+\frac{\alpha P}{MN_W}\right)=\frac{\alpha P}{N_0}\log e\;\;\mbox{[bits/sec]}.
\end{align*}

\begin{remark}\label{Remark2}
Under the assumption $N=8\alpha k^2 R_S^2$, the total received signal power in \eqref{eq:e36} is given by
\begin{align*}
E\left[\bfY^\dagger\bfY\right]
&=E\left[g^2\bfX^\dagger\bfSigma\bfPhi^\dagger\bfPhi\bfSigma\bfX\right]\\
&=E\left[8\alpha g^2 k^2 \bfX^\dagger\bfT^{\frac{1}{2}}\bftau\left(\frac{1}{N}\bfPhi^\dagger\bfPhi\right)\bftau\bfT^{\frac{1}{2}}\bfX\right]\\
&\rightarrow \alpha E\left[\bfX^\dagger\bfT\bfX\right]
\end{align*}
as $N\rightarrow\infty$, i.e., the received power tends to $\alpha$ times the transmit power.
\end{remark}

\begin{figure}[!t]
\centering
\includegraphics[angle=0,width=0.3\textwidth]{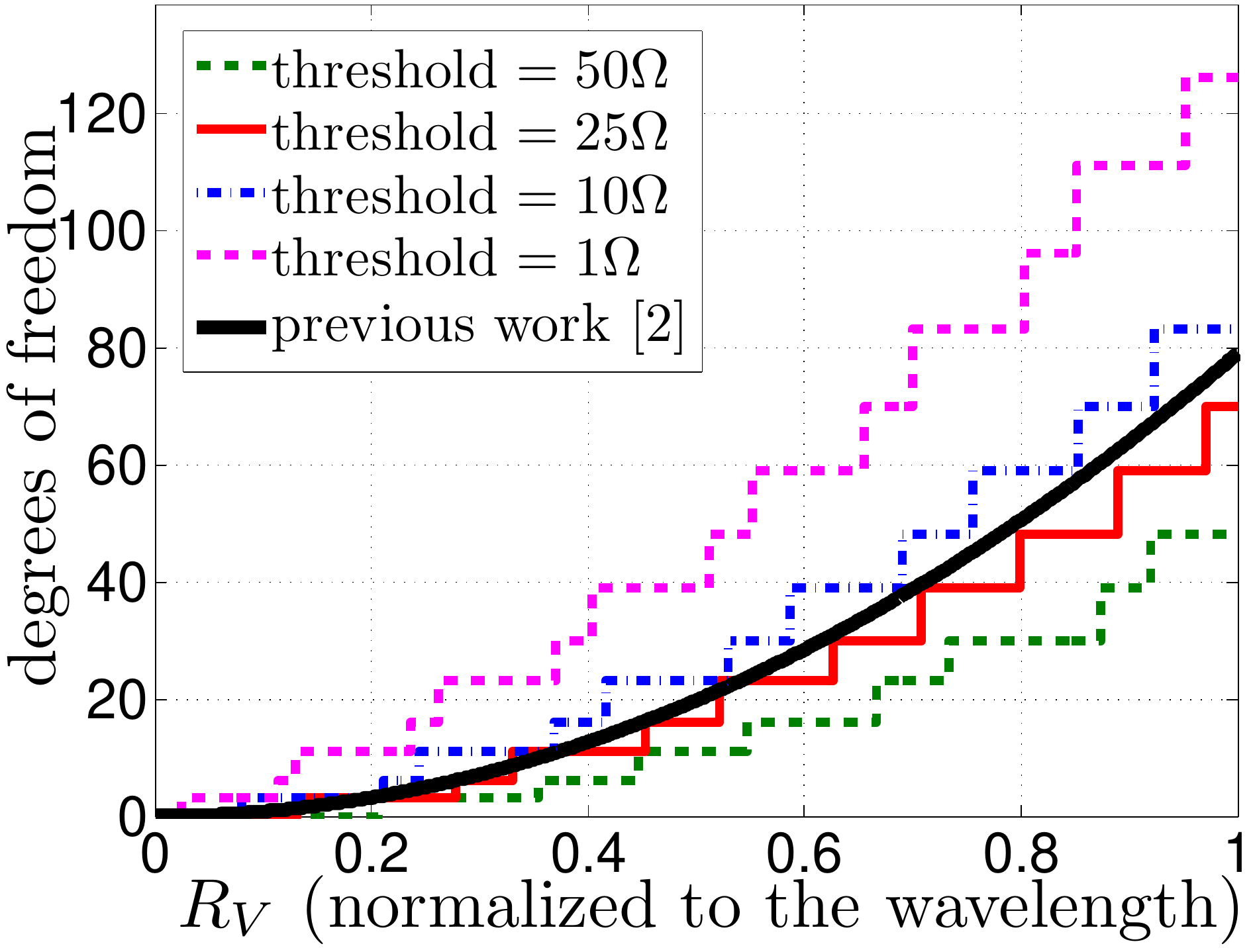}
\caption{Degrees of freedom with different thresholds}
\label{fig:comparison}
\end{figure}

\subsection{Comparison with ~\cite{PoonTse:2011}}
Our result shows the capacity of our channel is irrespective of the bandwidth and the size of the transmitter. This is because there are infinitely many decomposed channels of the same quality. This conclusion is different from that in~\cite{PoonTse:2011} that claimed the degrees of freedom is proportional to the surface area of $V$. The conclusions are different because in~\cite{PoonTse:2011} the number of useful channels is counted based on the singular values $\sigma_p$'s, which also vanishes as $n\rightarrow\infty$ similarly as the radiation resistance defined in \textbf{Definition \ref{sec:defi3}}. A small singular value or a small radiation resistance does not necessarily mean a less useful channel because it simply means we need to increase the amount of current to maintain the same transmit power. Thus, we actually have infinitely many useful channels. However, a small radiation resistance may be a problem for practical antenna design. If we count only the number of modes with radiation resistance bigger than a certain threshold, then our result is in line with that in~\cite{PoonTse:2011} as shown in Fig. \ref{fig:comparison} (when the threshold for the radiation resistance is between 10$\Omega$ and 25$\Omega$). Our result is a refinement to that in~\cite{PoonTse:2011} because our result shows explicitly how the degrees of freedom counted this way scales based on the threshold for the radiation resistance. Our result is practically important since it shows that, provided that a small radiation resistance is tolerable, even a compact antenna array with negligible bandwidth and antenna spacing well below the wavelength can provide a huge throughput as if the array was big enough so that the antenna spacing is on the order of the wavelength.

\appendices
\section{Proof of \textbf{Proposition \ref{Proposition1}}}\label{sec:AppA}
Due to limited space, we only show a proof sketch.
For notational simplicity, we use $h_{n}^{(1)},Y_{nm},P_{nm}$ to indicate $h_{n}^{(1)}(kr),Y_{nm}(\theta,\phi),P_{nm}(\cos\theta)$. 

Using \eqref{eq:E18}, we have 
\begin{align}
\notag\hfield(\bfr )
&=ik\sum_{p=1}^{\infty}\sigma_{p}J_{p}\curl\textbf{u}_{p}(\bfr ),\nonumber
\end{align}
where the Maxwell equation $\curl \efield=i\omega\mu\hfield$ is used where $\epsilon$ is the permittivity~\cite{Chew}. 
Then, the complex power flow leaving $S$ is
\begin{align*}
P_{c}
&\triangleq\oint_{S}\frac{1}{2}\left( \efield\times\hfield^{*}  \right)\cdot d \textbf{s}
=\frac{i \eta k^{3}}{2}
\sum_{p=1}^{\infty}\sum_{p'=1}^{\infty}
\sigma_{p} \sigma_{p'}^{*}J_{p}J_{p'}^{*}\gamma_{pp'},
\end{align*}
where 
$
\gamma_{pp'}\triangleq\oint_{S}\textbf{u}_{p}(\bfr )\times\curl\textbf{u}_{p'}^{*}(\bfr )\cdot d \textbf{s}
$
for all $p,p'\in\mathbb{Z}^+$.

\subsubsection{$l=l'=1$}\label{sec:case1}
\begin{align*}
\gamma_{pp'}
&=\frac{k}{C_p C_{p'}}\int\textbf{U}_{nm1}(\bfr)\times\textbf{U}_{n'm'2}^{*}\cdot \hat{\bfr}r^2 d\Omega\\
&=\frac{k}{C_p C_{p'}}\int\textbf{U}_{nm1}(\bfr)\cdot\textbf{U}_{n'm'2}^{*}\times\hat{\bfr}r^2 d\Omega\\
&=
\left\{
R_{S}+R_{S}^{2}\left.\left(\frac{1}{\hankfnp}\frac{d\hankfnp}{dr}\right)^{*}\right|_{r=R_{S}}
\right\}\delta_{pp'},
\end{align*}
which follows by vector identity $\left(\textbf{a}\times\textbf{b}\right)\cdot\textbf{c}=\textbf{a}\cdot\left(\textbf{b}\times\textbf{c}\right)$, (9.120) in \cite{Jackson}, and (40), (41), (42) and (44) in~\cite{PoonTse:2011}. 

\subsubsection{$l=1, l'=2$}\label{sec:case}
Similarly as in 1), we have
\begin{align}
\gamma_{pp'}
&\propto\oint_{S}
\left(
\nabla Y_{nm}\times \hat{\bfr}
\right)
\times
\left(
\nabla Y_{n'm'}^{*}\times \hat{\bfr}\right)
\cdot d\textbf{s}\nonumber\\
&\propto\int
\nabla Y_{n'm'}^{*}
\cdot
\left(
\hat{\bfr}\times \nabla Y_{nm}
\right) d \Omega\label{eq:ADD30}\\
&=\int
\hat{\bfr}\times\nabla Y_{n'm'}^{*}
\cdot
\hat{\bfr}\times\left(
\hat{\bfr}\times \nabla Y_{nm}
\right) d \Omega=0,\label{eq:ADD31}
\end{align}
where \eqref{eq:ADD30} follows by vector identity $\textbf{a}\times(\textbf{b}\times\textbf{c})=
(\textbf{a}\cdot\textbf{c})\textbf{b}-(\textbf{a}\cdot\textbf{b})\textbf{c}$, and \eqref{eq:ADD31} follows by (9.121) in \cite{Jackson}.

\subsubsection{$l=2,l'=1$}
$\gamma_{pp'}=0$ similarly as in 2).

\subsubsection{$l=l'=2$} The proof is similar to that in 1).
\begin{align}
\gamma_{pp'}
&=-\frac{C_{nm1}^2}{C_p^2}\left\{
R_{S}+R_{S}^{2}\left.\left(\frac{1}{\hankfn}\frac{d\hankfn}{dr}\right)\right|_{r=R_{S}}
\right\}\delta_{pp'}
\end{align}
Finally, we have
\begin{align*}
&\mathfrak{Re}
\left\{P_{c}\right\}
=\sum_{p=1}^{\infty}T_p
|J_{p}|^2,
\end{align*}
where
\begin{align*}
T_p\triangleq\frac{\eta k^3\sigma_p^2}{2}\mathfrak{Im}\left(-\gamma_{pp}\right)
\end{align*}
for $p\in \mathbb{Z}^+$. By using
\begin{align}
&\mathfrak{Im}
\left\{
\left.\left(\frac{1}{\hankfn}\frac{d\hankfn}{dr}\right)\right|_{r=R_{S}}
\right\}\nonumber=\frac{1}{k^2 R_S^2 |h_{n}^{(1)}(kR_{S})|^{2} },\nonumber
\end{align}
and $\sigma_p^2$ in \cite{PoonTse:2011}, we get $T_p$ in the proposition.
\section*{Acknowledgement}
This work was supported in part by the MSIP, Korea through the ICT R\&D Program 2013.

\bibliographystyle{IEEEtran}
\bibliography{IEEEabrv,References}

\begin{thebibliography}{10}
\providecommand{\url}[1]{#1}
\csname url@samestyle\endcsname
\providecommand{\newblock}{\relax}
\providecommand{\bibinfo}[2]{#2}
\providecommand{\BIBentrySTDinterwordspacing}{\spaceskip=0pt\relax}
\providecommand{\BIBentryALTinterwordstretchfactor}{4}
\providecommand{\BIBentryALTinterwordspacing}{\spaceskip=\fontdimen2\font plus
\BIBentryALTinterwordstretchfactor\fontdimen3\font minus
  \fontdimen4\font\relax}
\providecommand{\BIBforeignlanguage}[2]{{%
\expandafter\ifx\csname l@#1\endcsname\relax
\typeout{** WARNING: IEEEtran.bst: No hyphenation pattern has been}%
\typeout{** loaded for the language `#1'. Using the pattern for}%
\typeout{** the default language instead.}%
\else
\language=\csname l@#1\endcsname
\fi
#2}}
\providecommand{\BIBdecl}{\relax}
\BIBdecl

\bibitem{Franceschettietal:2009}
M.~Franceschetti, M.~D. Migliore, and P.~Minero, ``The capacity of wireless
  networks: information-theoretic and physical limits,'' \emph{{IEEE} Trans.
  Inf. Theory}, vol.~55, no.~8, pp. 3413--3424, Aug. 2009.

\bibitem{PoonTse:2011}
A.~S.~Y. Poon and D.~N.~C. Tse, ``Degree-of-freedom gain from using
  polarimetric antenna elements,'' \emph{{IEEE} Trans. Inf. Theory}, vol.~57,
  no.~9, pp. 5695--5709, Sep. 2011.

\bibitem{Andrewsetal:2001}
M.~R. {Andrews}, P.~P. {Mitra}, and R.~{deCarvalho}, ``Tripling the capacity of
  wireless communications using electromagnetic polarization,'' \emph{Nature},
  vol. 409, no.~8, pp. 316--318, Jan. 2001.

\bibitem{Chaeetal:2011}
S.~H. Chae, S.~W. Choi, and S.-Y. Chung, ``On the multiplexing gain of
  {$K$}-user line-of-sight interference channels,'' \emph{{IEEE} Trans.
  Commun.}, vol.~59, no.~10, pp. 2905--2915, Oct. 2011.

\bibitem{Gustafssonetal:2006}
M.~Gustafsson and S.~Nordebo, ``Characterization of {MIMO} antennas using
  spherical vector waves,'' \emph{{IEEE} Trans. Antennas Propagat.}, vol.~54,
  no.~9, pp. 2679--2682, Sep. 2006.

\bibitem{Getuetal:2005}
B.~N. Getu and J.~B. Andersen, ``The {MIMO} cube - a compact {MIMO} antenna,''
  \emph{{IEEE} Trans. Wireless Commun.}, vol.~4, no.~3, pp. 1136--1141, May
  2005.

\bibitem{Ivr:2011}
M.~T. Ivrla{$\check{\textup{c}}$} and J.~A. Nossek, ``Gaussian multiple access
  channel with compact antenna arrays,'' in \emph{Proc. {IEEE} International
  Symposium on Information Theory ({ISIT})}, Saint-Petersburg, Russia, 2011.

\bibitem{Chew}
W.~C. Chew, \emph{Waves and Fields in Inhomogeneous Media}.\hskip 1em plus
  0.5em minus 0.4em\relax New York:IEEE, 1995.

\bibitem{Balanis}
C.~A. Balanis, \emph{Antenna Theory}, 3rd~ed.\hskip 1em plus 0.5em minus
  0.4em\relax Wiley-Interscience, 2005.

\bibitem{Gradshteyn}
I.~S. Gradshteyn and I.~M. Ryzhik, \emph{Table of Integrals, Series, and
  Products}, 7th~ed.\hskip 1em plus 0.5em minus 0.4em\relax Academic Press,
  2007.

\bibitem{Jackson}
J.~D. Jackson, \emph{Classical Electrodynamics}, 3rd~ed.\hskip 1em plus 0.5em
  minus 0.4em\relax Hoboken, NJ:Wiley, 1998.

\end{thebibliography}

%
%

\end{document}